\documentclass{amsart}

\usepackage{amsmath}

\usepackage{amsthm}
\usepackage{amsfonts}
\usepackage{dsfont}
\usepackage{cite}

\newcommand{\reals}{\mathbb{R}}

\newcommand{\pb}[1]{\left\{#1\right\}}

\newcommand{\Ccomr}[3]{\big[#1,[#2,#3]\big]}

\newcommand{\para}[1]{\left(#1\right)}
\newcommand{\paraa}[1]{\big(#1\big)}
\newcommand{\parab}[1]{\Big(#1\Big)}
\newcommand{\parac}[1]{\bigg(#1\bigg)}

\theoremstyle{definition}

\theoremstyle{remark}

\numberwithin{equation}{section}

\newcommand{\vphi}{\varphi}
\newcommand{\eps}{\varepsilon}
\newcommand{\xv}{\vec{x}}

\renewcommand{\mid}{\mathds{1}}

\renewcommand{\d}{\partial}

\renewcommand{\d}{\partial}

\newcommand{\zt}{\tilde{z}}
\newcommand{\Zt}{\tilde{Z}}
\newcommand{\Wd}{W^\dagger}

\title[]{The world as quantized minimal surfaces}

\author{Joakim Arnlind}
\address[Joakim Arnlind]{Dept. of Math.\\
Link\"oping University\\
581 83 Link\"oping\\
Sweden}
\email{joakim.arnlind@liu.se}

\author{Jens Hoppe}
\address[Jens Hoppe]{Korean Institute for Advanced Study, 
Royal Institute of Technology,
Sogang University
}

\thanks{}

\subjclass[2000]{}
\keywords{}

\begin{document}

\begin{abstract}
  It is pointed out that the equations
  \begin{align*}
    \sum_{i=1}^d\Ccomr{X_i}{X_i}{X_j}=0
  \end{align*}
  (and its super-symmetrizations, playing a central role in M-theory
  matrix models) describe noncommutative minimal surfaces -- and can
  be solved as such.
\end{abstract}

\maketitle

\noindent During the past two decades several authors (see
e.g. \cite{M-Algebras,BFSS,IKKT,Cornalba}) have advocated the
equations
\begin{align}\label{eq:1}
  \sum_{i=1}^d\Ccomr{X_i}{X_i}{X_j}=0,
\end{align}
resp. the objects (specifically: self-adjoint infinite-dimensional
matrices) satisfying them as of potential relevance to understanding
space-time and the physical laws therein.

The analytical study of minimal surfaces on the other hand, going back
at least 250 years 
\cite{Lagrange,Meusnier,Euler} and being one of the most established
classical areas of mathematics, provides a wealth of explicit
examples, and very detailed knowledge of their properties (see
e.g. \cite{Nitsche,DHKW}).  In this note we would like to put forward
a direct relation between these two lines.

Parametrized minimal surfaces in Euclidean space are
solutions of $\Delta\xv=0$, where
\begin{align}\label{eq:2}
  \Delta := \frac{1}{\sqrt{g}}\d_a\sqrt{g}g^{ab}\d_b
\end{align}
is the Laplace operator on the embedded surface, and $g=\det(g_{ab})$ with
\begin{align}\label{eq:3}
  g_{ab}:=\sum_{i,j=1}^d
    \frac{\d x^i}{\d\vphi^a}\frac{\d x^j}{\d\varphi^b}\eta_{ij}
\end{align}
(here $\eta_{ij}=\delta_{ij}$ but one could equally well consider
general embedding spaces). Defining Poisson-brackets (with
$\rho=\rho(\vphi^1,\vphi^2)$)
\begin{align}
  \label{eq:4}
  \pb{f,h} := \frac{1}{\rho}\eps^{ab}\paraa{\d_af}\paraa{\d_b h}
\end{align}
the minimal surface equations can be written as
(cp. \cite{a:phdthesis,ahh:nambudiscrete,ah:dmsa})
\begin{align}
  \label{eq:5}
  \sum_{i=1}^d\pb{x_i,\pb{x_i,\xv}}-
  \frac{1}{2}\sum_{i=1}^d\frac{\rho^2}{g}\pb{x_i,g/\rho^2}\pb{x_i,\xv}=0,
\end{align}
hence as
\begin{align}
  \label{eq:6}
  \sum_{i=1}^d\pb{x_i,\pb{x_i,\xv}}=0
\end{align}
when choosing $\rho=±\sqrt{g}$, i.e.
\begin{align}
  \label{eq:7}
  \frac{g}{\rho^2}=\frac{1}{2}\sum_{i,j=1}^d\pb{x_i,x_j}^2=1.
\end{align}
While a general theory of non-commutative minimal surfaces, and
methods to construct them, will be given in a separate paper
\cite{ACH}, let us here focus on a particular example, the Catenoid,
\begin{align}
  \label{eq:8}
  \xv =
  \begin{pmatrix}
    \cosh v\cos u\\
    \cosh v\sin u\\
    v
  \end{pmatrix}=
  \begin{pmatrix}
    x \\ y \\ z
  \end{pmatrix}.
\end{align}
As $\xv_u^2=\xv_v^2=\cosh^2 v=\sqrt{g}$
\begin{align}
  \label{eq:9}
  \pb{x,y}=-\tanh z,\quad
  \pb{y,z}=\frac{x}{\cosh^2 z},\quad
  \pb{z,x}=\frac{y}{\cosh^2 z}.
\end{align}
One can easily verify \eqref{eq:6} , as well as (using
$x^2+y^2=\cosh^2z$) \eqref{eq:7}.

Following \cite{abhhs1,abhhs2} 
one could take e.g.
\begin{align}
  \label{eq:10}
  \begin{split}
  &[X,Y] = -i\hbar\tanh Z\\
  &[Y,Z] = (\cosh Z)^{-1}X(\cosh Z)^{-1}\\
  &[Z,X] = (\cosh Z)^{-1}Y(\cosh Z)^{-1}
  \end{split}
\end{align}
or (using power-series expansions for $(\cosh Z)^{-1}$) totally
symmetrized variants of \eqref{eq:10}, as defining a non-commutative
Catenoid. While it is easy to see that \eqref{eq:10} does have
solutions in terms of infinite-dimensional matrices $X,Y,Z$, it is
difficult to see whether or not these will satisfy \eqref{eq:1}. Let
us therefore first simplify the classical equations by defining
\begin{align}
  \label{eq:11}
  \zt(z):=\frac{z}{2}+\frac{1}{4}\sinh(2z),
\end{align}
satisfying
\begin{align}
  \label{eq:12}
  \frac{d\zt}{dz}=\cosh^2z>0
\end{align}
(hence being invertible, defining $z(\zt)$) as well as
\begin{align}
  \label{eq:13}
  \pb{x,y}=-t(\zt),\quad
  \pb{y,\zt}=x,\quad
  \pb{\zt,x}=y,
\end{align}
with $t(\zt):=\tanh z(\zt)$. The non-commutative analogue of
\eqref{eq:13},
\begin{align}
  \label{eq:14}
  [X,Y]=-i\hbar t(\Zt),\quad
  [Y,\Zt] =i\hbar X,\quad
  [\Zt,X] = i\hbar Y
\end{align}
resp. (defining $W=X+iY$)
\begin{align}
  \label{eq:17}
  [\Zt,W]=\hbar W,\quad
  [W,\Wd]=-2\hbar t(\Zt)
\end{align}
clearly has solutions where $\Zt$ is diagonal, with
\begin{align}
  \label{eq:18}
  \zt_j:=\Zt_{jj} = \zt_0-j\hbar=-j\hbar
\end{align}
and
\begin{align}
  \label{eq:19}
  W_{jk}=w_j\delta_{k,j+1};\quad
  |w_j|^2-|w_{j-1}|^2=-2\hbar t(-j\hbar).
\end{align}
When investigating \eqref{eq:1}, with
\begin{align}
  \label{eq:20}
  X_3=h(\Zt)=:H,\quad
  X_1+iX_2=W,
\end{align}
(the function $h$ to be determined) one finds that the two resulting
conditions (cp. (\ref{eq:1}))
\begin{align}
  \label{eq:21}
  \Ccomr{W}{\Wd}{H}=0
\end{align}
and
\begin{align}
  \label{eq:22}
  \frac{1}{2}\Ccomr{W}{\Wd}{W}+\Ccomr{H}{H}{W}=0
\end{align}
may be solved when deforming $[W,\Wd]$ to
\begin{align}
  \label{eq:23}
  \begin{split}
    &[W,\Wd]=-2\hbar T\\
    &T:=\tanh z(\Zt)+\hbar^2t_2(\Zt)+\sum_{n>2}^\infty\hbar^nt_n(\Zt),
  \end{split}
\end{align}
as well as taking the relation between $H$ and $\Zt$ to be of the form
\begin{align}
  \label{eq:24}
  H=z(\Zt)+\hbar^2h_2(\Zt)+\sum_{n>2}^\infty\hbar^nh_n(\Zt).
\end{align}
The advantage of keeping $[\Zt,W]=\hbar W$ undeformed is that then ($W$ still
being nonzero only on the first upper off-diagonal)
\begin{align}
  \label{eq:25}
  \begin{split}
    &f(\Zt)W = W f(\Zt+\hbar\mid)=:Wf_+\\
    &f(\Zt)\Wd = \Wd f(\Zt-\hbar\mid)=:\Wd f_-
  \end{split}
\end{align}
so that \eqref{eq:21} / \eqref{eq:22} can be seen to hold provided the
following finite-difference equations are satisfied:
\begin{align}
  &\hbar(T_+-T) = (H_+-H)^2\label{eq:26}\\
  &T\paraa{2H_+-H_{++}-H} = T_+\paraa{2H-H_+-H_-}\label{eq:27},
\end{align}
where $(H_{++})_{jj}=h_{++}(\Zt)_{jj}=h(\zt_j+2\hbar),\ldots$.
Assuming $T$ and $H$ to be monotonically increasing functions of $\Zt$
(and $\hbar>0$), one may write \eqref{eq:26} as
\begin{align}
  \label{eq:15}
  H_+-H=\sqrt{\hbar(T_+-T)},
\end{align}
which gives the condition
\begin{align}
  \label{eq:16}
  T\para{\sqrt{\frac{T_+-T}{T_{++}-T_+}}-1}
  =T_+\para{1-\sqrt{\frac{T_+-T}{T-T_-}}}
\end{align}
when inserting \eqref{eq:15} into \eqref{eq:27}. Using the expansion
for $T$ as given in \eqref{eq:23}, and Taylor-expanding
\begin{align}
  \label{eq:28}
  T_{\pm} = \tanh\paraa{z(\Zt\pm\hbar\mid)}+\hbar^2t_2(\Zt\pm\hbar\mid)+\cdots,
\end{align}
as well as $T_{++}$, one finds trivial agreement in $O(\hbar)$ while
the $\hbar^2$ resp. $\hbar^3$ terms demand
\begin{align}
  \label{eq:29}
  tt'''=\frac{3}{2}\frac{t(t'')^2}{t'}+t''t',
\end{align}
resp.
\begin{align}
  \label{eq:30}
  2t(t')^2t''''+6t(t'')^3-8tt't''t'''-3(t')^2(t'')^2=0 ;
\end{align}
using that for $t:=\tanh z(\zt)$ one has (with $c=c(\zt):=\cosh(z(\zt))$)
\begin{align}
  \label{eq:31}
  t'=\frac{1}{c^4},\quad
  t''=-\frac{4t}{c^6},\quad
  t'''=\frac{24}{c^8}-\frac{28}{c^{10}},\quad
  t''''=t\para{\frac{280}{c^{12}}-\frac{192}{c^{10}}}
\end{align}
it is straightforward to see that \eqref{eq:29} and \eqref{eq:30}
actually do hold (one should also note that in these orders $t_2$ does
not yet enter). Instead of deriving the 4th order expressions (which
give a third-order linear ODE for $t_2$), let us go back to \eqref{eq:15}
resp. \eqref{eq:26} which is consistently solved up to $O(\hbar^3)$
by $H=z(\Zt)$ and $T=t(\Zt)$, using
\begin{align}
  \label{eq:32}
  z'=\frac{1}{c^2},\quad
  z''=-\frac{2t}{c^4},\quad
  t' = (z')^2,\quad
  t''= 2z'z'' ,
\end{align}
while in order $\hbar^4$ giving the condition
\begin{align}
  \label{eq:33}
  t_2'-2z'h_2'=\frac{(z'')^2}{4}+\frac{2}{6}z'z'''-\frac{t'''}{6}
  =-\frac{1}{3}\frac{t^2}{c^8}
\end{align}
(using $z'''=\frac{8}{c^6}-\frac{10}{c^8}$, and \eqref{eq:32}). Both
$t_2$ (from \eqref{eq:16}, 4th order) and $h_2$ (from \eqref{eq:32})
are indeed small corrections to $t$, resp. $z$ (note that due to
$t'=1/c^4$, $c'=t/c$, any differential equation of the form
$f'=\frac{\alpha}{c^{n}}$ or $\frac{\alpha t}{c^{n}}$ can easily
be integrated), confirming the expectation that the power-series in
\eqref{eq:23} and \eqref{eq:24} actually make sense (as formal
power-series or asymptotic series, or even as series actually converging for small $\hbar$; note
that due to the unboundedness of the eigenvalues of $\Zt$ it is
necessary that $h_2(\zt_j)$ and $t_2(\zt_j)$ are small corrections to
$z_j=z(\zt_j)$ resp. $t(\zt_j)$ for all $j$).

In accordance with the classical Casimir relation
\begin{align}
  \label{eq:34}
  x^2+y^2-\cosh^2z(\zt)=x^2+y^2-c^2=0
\end{align}
one may also look for $E=e(\Zt)$ such that
\begin{align}
  \label{eq:35}
  \frac{1}{2}\paraa{W\Wd+\Wd W} = E =
  c^2+\sum_{n\geq 2}\hbar^ne_n(\Zt).
\end{align}
The condition (take the commutator of \eqref{eq:35} with $W$, using
\eqref{eq:23})
\begin{align}
  \label{eq:36}
  \begin{split}
    0 &= \hbar\paraa{WT+TW}-[E,W]\\
    &= W\paraa{\hbar T+\hbar T_++E-E_+}
  \end{split}
\end{align}
necessitates
\begin{align}
  \label{eq:37}
  \hbar
  e_0'+\hbar^2\frac{e_0''}{2}+\hbar^3\parac{\frac{e_0'''}{6}+e_2'}
  =\hbar 2t+\hbar^2t'+\hbar^3\para{\frac{t''}{2}+2t_2}
\end{align}
i.e. (using $e_0=c^2$, $e_0'=2cc'=2t$, $e_0^{(n)}=2t^{(n-1)}$)
\begin{align}
  \label{eq:38}
  e_2'=\frac{t''}{6}+2t_2=2t_2-\frac{2t}{3c^6}.
\end{align}
As a consistency-check consider again (\ref{eq:21}), yielding
\begin{eqnarray}
\label{eq:39} WW^{\dagger}&=& 2\hbar\frac{H_{+}-H}{2H-H_{+}-H_{-}}T\\
\label{eq:40} W^{\dagger}W&=&2\hbar\frac{H-H_{-}}{2H-H_{+}-H_{-}}T\,,
\end{eqnarray}
but then using (\ref{eq:35}), resulting in
\begin{equation}\label{eq:41}
\hbar(H_{+}-H_{-})T=E(2H-H_{+}-H_{-})\,,
\end{equation}
which is consistently solved in\,\,$O(\hbar^{2})$ and $O(\hbar^{3})$
while requiring
\begin{equation}\label{eq:42}
  c^{2}h''_{2}-\frac{2t}{c^{4}}e_{2}+\frac{2}{c^{2}}t_{2}+2th_{2}'
  =\frac{t}{3}\para{\frac{10}{c^{8}}-\frac{8}{c^{6}}}
  -\frac{c^{2}z''''}{12} 
  = \frac{t}{3}\para{\frac{4}{c^{6}}-\frac{10}{c^{8}}}
\end{equation}
when comparing terms proportional to $\hbar^{4}$.

Using (\ref{eq:38}) and (\ref{eq:33}), as well as
$z''''=\frac{-48t}{c^{8}}+\frac{80t}{c^{10}}$, then yields a 3rd order
ODE for $ e_{2}$, (just as if inserting (\ref{eq:38}) and
(\ref{eq:33}) into the third-order ODE for $t_{2}$ that results in 4th
order from (\ref{eq:16}))\ ,
\begin{equation}\label{eq:44}
  \frac{c^{4}}{4}e_{2}'''+tc^{2}e_{2}''+\frac{e_{2}'}{c^{2}}
  -\frac{2t}{c^{4}}e_{2}=2t\para{-\frac{1}{c^{6}}+\frac{1}{c^{8}}},
\end{equation}
which is in fact slightly simpler than the one for $t_{2}$,
\begin{equation}\label{eq:45}
  \frac{tc^{12}}{2}t_{2}'''+t_{2}''\para{6c^{10}-\frac{13}{2}c^{8}}+
  tt_{2}'\para{12c^{8}-10c^{6}}-2c^{2}t_{2}+t\para{\frac{16}{c^{4}}
  -\frac{20}{c^{2}}+4}=0
\end{equation}
that follows from (\ref{eq:33})/(\ref{eq:38})/(\ref{eq:42}) (and is
identical to the $\hbar^{4}$-condition following from (\ref{eq:16})).
Taking
\begin{equation}\label{eq:46}
  e_{2}=\frac{1}{18}\para{4-\frac{2}{c^{2}}+\frac{1}{c^{4}}}
\end{equation}
as a solution of (\ref{eq:44}) one finds / can choose
\begin{equation}\label{eq:47}
  t_{2}=\frac{t}{9}\para{\frac{1}{c^{4}}+\frac{2}{c^{6}}},
  \quad h_{2}=\frac{t}{90}\para{-4+\frac{8}{c^{2}}+\frac{11}{c^{4}}}.
\end{equation}
Note that $t_{2}$ and $h_{2}$ (both odd) and $e_{2}$ (even) are indeed
small corrections to $t(\tilde{Z})=\tanh z(\tilde{Z})$ and
$z(\tilde{Z})$ (resp. $c^{2}=\cosh^2 z(\tilde{Z})$) consistent
with our claim that (\ref{eq:23})/(\ref{eq:24})
resp. (\ref{eq:18})/(\ref{eq:19})/(\ref{eq:20}) (with $t$ replaced by
$T$) define solutions of (\ref{eq:1}), which for $\hbar \rightarrow 0$
converge to the classical commutative catenoid (described by Euler in
1744 \cite{Euler}). 

Let us comment that (cp. (\ref{eq:7}))
\begin{equation}\label{eq:48}
G:=-\frac{1}{\hbar^{2}}\sum_{i<j}\lbrack X_{i}, X_{j}\rbrack^{2}
\end{equation}
is indeed equal to $\mid$ to leading order (though not to all
orders):
\begin{align}
  \label{eq:49}
  \begin{split}
    &\!\!\!\hbar^{2}G =\frac{1}{2}\para{\lbrack H, W\rbrack
    \lbrack W^{\dagger}, H\rbrack+\lbrack W^{\dagger}, H\rbrack
    \lbrack H,W \rbrack}-\lbrack X, Y\rbrack^{2}\\
    &=\frac{1}{2}\para{(H-H_{-})WW^{\dagger}(H-H_{-})
    +(H_{+}-H)W^{\dagger}W(H_{+}-H)}-\lbrack X, Y\rbrack^{2}\\
    &=\hbar T\para{\frac{(H-H_{-})^{2}(H_{+}-H)}{2H-H_{+}-H_{-}}
    +\frac{(H_{+}-H)^{2}(H-H_{-})}{2H-H_{+}-H_{-}}}-
    \lbrack X,Y\rbrack^{2}\\
    &=\hbar T(H_{+}-H)(H-H_{-})\para{\frac{H_{+}-H_{-}}{2H-H_{+}-H_{-}}
    }-\lbrack X, Y\rbrack^{2}\\
    &=(H_{+}-H)(H-H_{-})E+\hbar^{2}T^{2}\\
    &=\hbar^{2}\bigg(\para{(z')^{2}+\hbar^{2}\para{\frac{z'z'''}{3}+
        2h_{2}z'-\frac{(z'')^{2}}{4}}+\cdots}(c^{2}+\hbar^{2}e_{2}+\cdots)\\
    &\qquad+(t+\hbar^{2}t_{2}+\cdots)^{2}\bigg);    
  \end{split}
\end{align}
while in leading order one thus gets
\begin{equation}\label{eq:50}
G_{0}=(z')^{2}c^{2}+t^{2}=\frac{1}{c^{2}}+t^{2}=\mid,
\end{equation}
the terms proportional to $\hbar^{2}$,
\begin{equation}\label{eq:51}
  (z')^{2}e_{2}+c^{2}\para{\frac{z'z'''}{3}+2h_{2}z'
  -\frac{(z'')^{2}}{4}}+2tt_{2}=\frac{1}{18}\para{\frac{40}{c^{6}}
  -\frac{43}{c^{8}}}
\end{equation}
do not cancel, but are bounded $(\in\lbrack-\frac{1}{6},\frac{1}{4}))$
and because of $\hbar^{2}$ therefore small correction to $\mid$.

Note that due to the commutation relation (cp.(\ref{eq:23}))
\begin{align}\label{eq:52}
  \begin{split}
    [X_{1}, X_{2}] &= -i\hbar T\\    
    [\tilde{Z}, X_{1}+iX_{2}]&=\hbar(X_{1}+iX_{2}),
  \end{split}
\end{align}
with $T\approx \tilde{Z}$ near the ''middle'' of the infinite
dimensional matrix (where, due to $(\cosh z(\tilde{Z}))^{2}\approx
\mid+\tilde{Z}^{2}$, $X_{1}^{2}+X_{2}^{2}-X_{3}^{2}\approx
\mid$) one also could think of the non-commutative catenoid as a
particular infinite dimensional `unitarizable' representation of a
non-linear deformation of $so(2,1)$.

Let us summarize: we have shown how to construct 3
infinite-dimensional matrices $X_{i}$ $(i=1,2,3)$, correponding to the
embedding functions of the classical catenoid in $\reals^{3}$,
satisfying
\begin{equation}\label{eq:53}
  \sum_{i=1}^{3}\Ccomr{X_i}{X_i}{X_j}=0,
\end{equation}
explicitely checked up to several orders in $\hbar$.  Concretely,
\begin{align}\label{eq:54}
  \begin{split}
    &(X_{3})_{jk}=\delta_{jk}\parab{z_{j}+\hbar^{2}\frac{t_{j}}{90}
    \parab{-4+\frac{8}{c_{j}^{2}}+\frac{11}{c_{j}^{4}}}+\cdots}\\    
    &(X_{1}+iX_{2})_{jk}=w_{j}\delta_{k,j+1}\\
    &\vert w_{j}\vert^{2}-\vert w_{j-1} \vert^{2}=
    -2\hbar t_{j}\parab{1+\frac{\hbar^{2}}{9}\parab{\frac{1}{c_{j}^{4}}
    +\frac{2}{c_{j}^{6}}}+\cdots}
  \end{split}
\end{align} 
where (cp. (\ref{eq:11})) $\tilde{z}_{j}=-j\hbar$, 
$z_{j}=z(\tilde{z}_{j})$, $t_{j}=\tanh z(\tilde{z}_{j})$, 
$c_{j}=\cosh z(\tilde{z_{j}})$.

\section*{Acknowledgment}

\noindent We thank Jaigyoung Choe for collaboration (on a general theory
of noncommutative minimal surfaces), and Ki-Myeong Lee for a
discussion concerning the IKKT model.


\begin{thebibliography}{AA}

\bibitem{M-Algebras}
J.Hoppe.
\newblock On M-Algebras, the Quantisation of Nambu-Mechanics, and Volume Preserving Diffeomorphisms,  
\newblock hep-th/9602020,  {\em Helv.Phys.Acta} 70 (1997) 302-317

\bibitem{BFSS}
T.Banks, W.Fischler, S.Shenker, L.Susskind.
\newblock 
M Theory As A Matrix Model: A Conjecture, 
\newblock hep-th/9605168, {\em Phys.Rev.} D55:5112-5128,1997

\bibitem{IKKT}
N.Ishibashi, H.Kawai, Y.Kitazawa, A.Tsuchiya.
\newblock A Large-N Reduced Model as Superstring,
\newblock hep-th/9612115, {\em Nucl.Phys.} B498 (1997) 467-491

\bibitem{Cornalba}
L.Cornalba, W.Taylor.
\newblock Holomorphic curves from matrices,
\newblock hep-th/9807060 {\em Nucl.Phys.} 536:513-552,1998

\bibitem{Lagrange}
J.L.Lagrange.
Essai d'une nouvelle methode pour determiner les maxima et les minima des formules integrales
indefinies. Miscellanea Taurinensia 2, 173-195 (1760-1762). Oeuvres, vol. I. Gauthier-Villars,
Paris 1867, pp. 335-362

\bibitem{Meusnier}
J.B.Meusnier.
Memoire sur la courbure des surfaces. Memoire des savants strangers 10 (lu 1776), 477-510 (1785)

\bibitem{Euler}
L. Euler. Methodus inveniendi lineas curvas maximi minimive proprietate gaudentes, 1744, in: Opera omnia I, 24

\bibitem{Nitsche}
J.C.C.Nitsche. 
Vorlesungen \"uber Minimalfl\"achen, Die Grundlehren der
mathematischen Wissenschaften in Einzeldarstellungen, Band 199,
Springer-Verlag, Berlin Heidelberg, New York, 1975.


\bibitem{DHKW} U.Dierkes, S.Hildebrandt, A.K\"uster, O.Wohlrab.
  Minimal surfaces I. Springer-Verlag, Berlin Heidelberg, New York,
  1992.

\bibitem{a:phdthesis}
J.Arnlind.
\newblock {\em Graph Techniques for Matrix Equations and Eigenvalue Dynamics}.
\newblock PhD thesis, Royal Institute of Technology, 2008.

\bibitem{ah:dmsa}
J.Arnlind, J.Hoppe.
\newblock Discrete minimal surface algebras.
\newblock {\em SIGMA Symmetry Integrability Geom. Methods Appl.}, 6, 2010.


\bibitem{ahh:nambudiscrete}
J.~Arnlind, J.~Hoppe, G.~Huisken.
\newblock Multi-linear formulation of differential geometry and matrix
  regularizations.
\newblock {\em J. Diff. Geo.}, 91:1--39, 2012.

\bibitem{ACH}
J.Arnlind, J.Choe, J.Hoppe.
\newblock Noncommutative Minimal Surfaces \textit{(in preparation)}.

\bibitem{abhhs1}
J.Arnlind, M.Bordemann, L.Hofer, J.Hoppe, 
H.Shimada;
\newblock Fuzzy {R}iemann Surfaces, 
\newblock {\em J. High Energy Phys.}, JHEP06(2009)047. 

\bibitem{abhhs2}
J.Arnlind, M.Bordemann, L.Hofer, J.Hoppe, H.Shimada.
\newblock Noncommutative {R}iemann surfaces by embeddings in {$\Bbb R\sp
3$}.
\newblock {\em Comm. Math. Phys.}, 288(2):403--429, 2009.

\end{thebibliography}
\end{document}